# An Improved Active Disturbance Rejection Control for a Differential Drive Mobile Robot with Mismatched Disturbances and Uncertainties

Ibraheem Kasim Ibraheem
Electrical Engineering Department
College of Engineering, Baghdad University
Baghdad, Iraq
ibraheemki@coeng.uobaghdad.edu.iq

Wameedh Riyadh Abdul-Adeem
Electrical Engineering Department
College of Engineering, Baghdad University
Baghdad, Iraq
Wameedh.r@coeng.uobaghdad.edu.iq

*Abstract*—In this paper a new strategy based on disturbance and uncertainty (DU) estimation and attenuation technique is proposed and tested on the nonlinear kinematic model of the differential drive mobile robot (DDMR). The proposed technique is an improved version of the Active Disturbance Rejection Control (ADRC) strategy suggested by J. Han. The ADRC is used to actively reject disturbances caused by the unknown exogenous signals and the matched uncertainties of the system model, which are lumped all together and attributed as a total disturbance. In this work, the considered system is assumed to be affine and the total disturbance and the input are considered to be on different channels. To deal with the mismatched disturbances and uncertainties, the total disturbance has been converted into a matched one. Then, based on the improved ADRC (IADRC), the dynamic performance of the DDMR has been enhanced by estimating the total disturbance and canceling it from the system. Through digital simulations, different performance measures are applied, and they all indicate the effectiveness of the proposed IADRC by almost removing the chattering phenomenon and providing a high immunity in the closed-loop system against torque disturbance.

*Keywords*—Improved active disturbance rejection control; total disturbances; Differential drive mobile robot; orientation error

## I. INTRODUCTION

The DU broadly presented in most of the engineering applications and attend opposite influences on the performance of control systems [1]. The attenuation of the DU is a strategic objective in control engineering. When a disturbance is detectable through measurement, a feed-forward method could attenuate or reject the effect of disturbance [2]. However, the exogenous disturbance cannot be measured or is exceptionally expensive to measure. The first spontaneous thought to treat with this challenging is to build an observer to estimate the disturbance. Then, an action signal can be established to compensate the effect of the disturbance. This simple indication can be broadened to reject the uncertainties. The effect of the uncertainties or the dynamics that are not modeled could be estimated as a part of the disturbance. So, a new term of disturbance acted, i.e. the "total disturbance," which describes the aggregation of the input disturbances and system uncertainties. This class of techniques is denoted as Estimation and Attenuation of Disturbance/Uncertainty (EAD/U). A several EAD/U structures were individually suggested, Han firstly suggested an Extended State Observer (ESO) in the 1990s [3]. An ESO is mostly viewed as an essential role of the technique termed active disturbance rejection control (ADRC) [4]. The ADRC consists of three essential parts: a tracking differentiator, an extended state observer, and a nonlinear state error feedback controller to solve the control problems in various applications with promising results.

The ADRC as a complete structure has been utilized in the real manufacturing appliances; the ADRC method has been used to achieve the high-precision control of ball screw feed drives [5]. Also, a dual-loop ADRC algorithm that is used for an active hydraulic suspension system, which can help the six-wheel off-road vehicle to improve the performance transition [6]. In the field of robots, the ADRC is useful in quad helicopter control due to superiority to solve control problems and disturbance estimation of the nonlinear models with uncertainty and intense disturbances superiority [7]. Additionally, the success has been established by many engineering systems[8-10]. The main objective of this paper is to design a controller which provides an active rejection of the bounded mismatched total disturbances which has direct effect on the performance of PMDC motors of the DDMR. The controller guarantees a minimum orientation error in spite of disturbances. The disturbance includes friction torques, fluctuations of the load, change of parameters for the actuators, and external disturbance occurs due to a collision with obstacles.

The contribution of this paper lies in applying an improved version of the classical Han's ADRC in the motion control of the DDMR, which is a nonlinear MIMO system, It is an extension of our three previous published papers [11, 12, 13]. The proposed IADRC is constructed by combining three primary units. The first unit is the improved nonlinear tracking differentiator (INTD), which is used to extract differentiation of any piecewise smooth nonlinear signal to reach a high accuracy. Also, the INTD attenuates signals with frequencies outside a certain frequency band. The Improved nonlinear state error feedback (INSEF) controller is the second unit in the proposed controller. This unit is derived by combining the nonlinear gains and the PID controller with a new control structure. The last unit is the sliding mode extended state observer (SMESO), it is an extension of the LESO method; it performs better than the LESO observer in terms chattering reduction in the control signal by including a nonlinearity and a sliding mode term on the LESO.

The rest of the paper is organized as follows: Section II presents the improved ADRC. The mathematical models of DDMR and PMDC are introduced in section III. Section IV illustrates the numerical simulations and some comments and highlights on the work. Finally, conclusions are given in section V.

## II. THE IMPROVED ACTIVE DISTURBANCE REJECTION CONTROL (IADRC)

The classical ADRC proposed by J. Han is built by combining the tracking differentiator (TD), the nonlinear state error combination (NLSEF), and the linear extended state observer (LESO); the entire structure presented in[4, 14].

In fig. 1, the improved version of the ADRC is illustrated. The following subsections discuss each part of the proposed controller supported by necessary explanations.

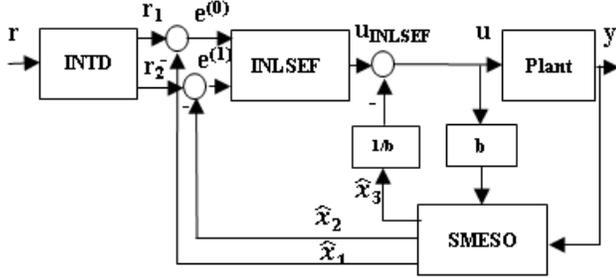

Fig.1. The IADRC topology

### A. The Improved Nonlinear Tracking Differentiator

Classical Han's tracking differentiator has been improved be using a smooth nonlinear function tanh(.) instead of the sign(.) function. The hyperbolic tangent tanh(.) function is introduced due to two reasons. Firstly, the slope of the continuous function tanh(.) near the origin significantly accelerates the convergence of the proposed tracking differentiator and reduces the chattering phenomenon. Secondly, the saturation feature of the function due to its nonlinearity increases the robustness against noise signals. A second improvement is accomplished by combining both the linear and the nonlinear terms. This tracking differentiator shows a better dynamical performance than Han's tracking differentiator.

The improved nonlinear tracking differentiator has been designed based on the hyperbolic tangent function [11]:

$$\left.\begin{array}{l}\dot{r}_1 = r_2 \\ \dot{r}_2 = -R^2 \tanh\left(\frac{\beta r_1 - (1-\alpha)r}{\gamma}\right) - R r_2\end{array}\right\} \quad (1)$$

Where $r_1$ is tracking the input r, and $r_2$ tracking the derivative of input r. the parameters $\alpha, \beta, \gamma,$ and R are appropriate design parameters, where $0 < \alpha < 1, \beta > 1, \gamma > 0,$ and $R > 0$.

The configuration with the proposed INTD can eliminate the chattering phenomenon and measurement noise effectively with fast arrival and smooth tracking to the input signal. The stability of the suggested tracking differentiator is proven based on a proposed Lyapunov function in [11].

### B. The Improved Nonlinear State Error Feedback Controller

The original version of the nonlinear feedback functions in the forms of *fal*(.) was firstly proposed by Han [4], which is continuous and smooth. The improved nonlinear state error feedback control law provides more shape flexibility on a wide range of the state error vector. This behavior improves both the performance and the robustness of the controlled system. The improved nonlinear algorithm using the *sign*(.) and exponential functions is established as follows[12]:

$$u_{INLSEF} = \Psi(e) = k(e)^T f(e) + u_{integrator} \quad (2)$$

Where **e** is n × 1 state error vector, defined as:
$$\mathbf{e} = [e^{(0)} \quad \ldots e^{(i)} \ldots \quad e^{(n-1)}]^T$$
$e^{(i)}$ is the *n-th* derivative of the state error, defined as:
$$e^{(i)} = r_{i+1} - \hat{x}_{i+1}$$

$k(e)$ is the nonlinear gain function, defined as:

$$k(\mathbf{e}) = \begin{bmatrix} k(\mathbf{e})_1 \\ \vdots \\ k(\mathbf{e})_i \\ \vdots \\ k(\mathbf{e})_n \end{bmatrix} = \begin{bmatrix} \left(k_{11} + \frac{k_{12}}{1+exp(\mu_1(e^{(0)})^2)}\right) \\ \vdots \\ \left(k_{i1} + \frac{k_{i2}}{1+exp(\mu_n(e^{(i-1)})^2)}\right) \\ \vdots \\ \left(k_{n1} + \frac{k_{n2}}{1+exp(\mu_n(e^{(n-1)})^2)}\right) \end{bmatrix} \quad (3)$$

The coefficients $k_{i1}$, $k_{i2}$, and $\mu_i$ are positive constants. The advantage of the nonlinear gain term $k(e)_i$ is to make the nonlinear controller much more sensitive to small values. When $e^{(i-1)} = 0$, $k(e)_i = k_{i1} + k_{i2}/2$, while as $e^{(i-1)}$ goes large enough $k(e)_i \approx k_{i1}$. For values of $e^{(i-1)}$ in between, The nonlinear gain $k(e)_i$ term is bounded in the sector [ $k_{i1}$, $k_{i1}+k_{i2}/2$]. The function $f(e)$ is the error function, defined as:

$$f(\mathbf{e}) = \left[|e^{(0)}|^{\alpha_1}sign(e) \; \ldots |e^{(i)}|^{\alpha_i}sign(e^{(i)}) \ldots \; |e^{(n-1)}|^{\alpha_n}sign(e^{(n)})\right]^T \quad (4)$$

Eq. (4) shows significant features in the nonlinear term $|e|^{\alpha_i}$. For $\alpha_i \ll 1$, the term $|e|^{\alpha_1}$ is rapidly switching. This feature makes the error function $f(e)$ is sensitive for small error values. As $\alpha_i$ goes beyond 1, the nonlinear term becomes less sensitive for small variations in the error signal *e*.

### C. Sliding Mode Extended State Observer

The proposed SMESO has the following state space representation [13]:

$$\dot{\hat{X}} = F\hat{X} + B_1 u + B_2 g(y - \hat{x}_1) \quad (5)$$

where $\hat{X} \in R^{(n+1)\times 1}$, is a vector that contains the estimated plant states and the total disturbance, $\dot{\hat{X}} \in R^{(n+1)\times 1}, B_1 \in R^{(n+1)\times 1}, B_2 \in R^{(n+1)\times 1}, F \in R^{(n+1)\times(n+1)}$.

$$\hat{X} = [\hat{x}_1 \quad \hat{x}_2 \; \ldots \; \hat{x}_{n+1}]^T, \quad \dot{\hat{X}} = [\dot{\hat{x}}_1 \quad \dot{\hat{x}}_2 \; \ldots \; \dot{\hat{x}}_{n+1}]^T$$

$$F = \begin{bmatrix} 0 & 1 & 0 & 0 & \cdots & 0 \\ 0 & 0 & 1 & 0 & \cdots & 0 \\ 0 & 0 & 0 & 1 & \cdots & 0 \\ 0 & \vdots & \vdots & \vdots & \ddots & \vdots \\ 0 & 0 & 0 & 0 & \cdots & 1 \\ 0 & 0 & 0 & 0 & 0 & 0 \end{bmatrix}$$

$$B_1 = [0 \quad 0 \; \ldots 1 \quad 0]^T, \quad B_2 = [\beta_1 \quad \beta_2 \; \ldots \; \beta_{n+1}]^T$$

Now, $g(y - \hat{x}_1) = K_\alpha |y - \hat{x}_1|^\alpha sign(y - \hat{x}_1) + K_\beta |y - \hat{x}_1|^\beta (y - \hat{x}_1)$

Where $K_\alpha$, $\alpha$, $K_\beta$, and $\beta$ are appropriate design parameters. For $n = 2$, the nonlinear state space representation of the proposed SMESO is given as:

$$\left.\begin{array}{l}\dot{\hat{x}}_1 = x_2 + \beta_1(K_\alpha|y - \hat{x}_1|^\alpha sign(y - \hat{x}_1) \\ \qquad + K_\beta|y - \hat{x}_1|^\beta(y - \hat{x}_1)) \\ \dot{\hat{x}}_2 = x_3 + bu + \beta_2(K_\alpha|y - \hat{x}_1|^\alpha sign(y - \hat{x}_1) \\ \qquad + K_\beta|y - \hat{x}_1|^\beta(y - \hat{x}_1)) \\ \dot{\hat{x}}_3 = \beta_3(K_\alpha|y - \hat{x}_1|^\alpha sign(y - \hat{x}_1) \\ \qquad + K_\beta|y - \hat{x}_1|^\beta(y - \hat{x}_1))\end{array}\right\} \quad (6)$$

The SMESO is an extension of the LESO method, which as a state estimator it performs better than the LESO observer in terms of chattering reduction in the control signal. It was proven in [13] that the estimation error is asymptotically convergent to zero under certain conditions in the nonlinear gain function. The estimation accuracy has been increased by adding the sliding term in the nonlinear extended state observer. The proposed method achieves an outstanding performance in terms of smoothness in the control signal which means that less control energy is required to achieve the desired performance [13].

### D. Handling the mismatched total disturbance

The IADRC (Eq. (1)-Eq. (3)) uses the same structure of Han's one. This structure has been utilized to reject the matched total disturbance. Therefore, to deal with the mismatched total disturbance, it needs to be transformed into matched one. The following proposed theorem transforms the mismatched total disturbance in Eq. (7) in to matched one given by Eq. (11).

**Theorem 1**

Consider the second-order affine nonlinear dynamical system with mismatched bounded disturbance that is represented by:

$$\left. \begin{array}{l} \dot{x}_1 = f_1(x_1, x_2) + b_1 d \\ \dot{x}_2 = f_2(x_1, x_2) + b_2 u \\ y = x_1 \end{array} \right\} \quad (7)$$

This system can be transformed into the model with the state space given by:
$$\dot{x}_1 = x_3$$
$$\dot{x}_3 = \hat{f}(x_1, x_2) + \hat{b}(u + \hat{d})$$
$$y = x_1$$
where
$$\hat{f}(x_1, x_2) = \frac{\partial f_1(x_1, x_2)}{\partial x_1} f_1(x_1, x_2) + \frac{\partial f_1(x_1, x_2)}{\partial x_2} f_2(x_1, x_2)$$
$$\hat{b} = b_2 \frac{\partial f_1(x_1, x_2)}{\partial x_2}$$
$$\hat{d} = (b_1 \frac{\partial f_1(x_1, x_2)}{\partial x_1} d + b_1 \dot{d}) / (b_2 \frac{\partial f_1(x_1, x_2)}{\partial x_2})$$

**proof**

Differentiate (7a) with respect to $t$, then:
$$\ddot{x}_1 = \frac{\partial f_1(x_1, x_2)}{\partial x_1} \dot{x}_1 + \frac{\partial f_1(x_1, x_2)}{\partial x_2} \dot{x}_2 + b_1 \dot{d} \quad (8)$$

Substitute (7) into (8) to get
$$\ddot{x}_1 = \frac{\partial f_1(x_1, x_2)}{\partial x_1} (f_1(x_1, x_2) + b_1 d) + \frac{\partial f_1(x_1, x_2)}{\partial x_2} (f_2(x_1, x_2) + b_2 u) + b_1 \dot{d} \quad (9)$$

Rearrange (9), then:
$$\ddot{x}_1 = \frac{\partial f_1(x_1, x_2)}{\partial x_1} f_1(x_1, x_2) + \frac{\partial f_1(x_1, x_2)}{\partial x_2} f_2(x_1, x_2) + b_2 \frac{\partial f_1(x_1, x_2)}{\partial x_2} u + b_1 \dot{d} + b_1 \frac{\partial f_1(x_1, x_2)}{\partial x_1} d \quad (10)$$

Then (10) reduces to
$$\ddot{x}_1 = \hat{f}(x_1, x_2) + \hat{b}(u + \hat{d})$$
Let, $x_3 = \dot{x}_1, \xi = x_2$
Then
$$\left. \begin{array}{l} \dot{x}_1 = x_3 \\ \dot{x}_3 = \hat{f}(x_1, \xi) + \hat{b}(u + \hat{d}) \\ \dot{\xi} = f_2(x_1, \xi) + b_2 u \end{array} \right\} \quad (11)$$

Finally, (11) is called the canonical form of ADRC [15]. □

## III. MODELING OF DIFFERENTIAL DRIVE MOBILE ROBOT

The mobile robot mathematical model is an approximation of the physical mobile robot which consists of the kinematic and actuators dynamical models. To restrain the robot's motor dynamics, an internal loop is also involved – the block scheme of the mobile robot is showed on fig. 2 [16].

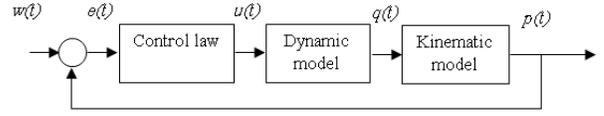

Fig.2 Mobile robot with internal control loop

The mobile robot's reference input $w(t)$ are the required velocities, the output of internal control loop $q(t)$ are the robot's recent velocities, which are later altered to produce the robot posture $p(t)$ by the kinematic model. The control inputs are the differences between the required and the recent velocities $e(t) = w(t) - q(t)$ while the control output $u(t)$ influences the dynamics of the mobile robot as forces or torques. The posture of the mobile robot in regard to the origin of the global coordinate system (GCS) is described by the position coordinates $x, y$ of its local coordinate system (LCS) origin with rotation defined by an angle $\theta_m$ [16].

As seen in fig. 3, the kinematic model can be described by the robot's linear velocity $V_m$ and its angular velocity $\omega_m$. But for the most control configurations; it is desirable to describe it by the wheel angular velocities $\omega_{wr}, \omega_{wl}$. The general kinematic model of DDMR defined as [17-20]:
$$\dot{x}' = V_m cos(\theta_m)$$
$$\dot{y}' = V_m sin(\theta_m) \quad (12)$$
$$\dot{\theta}_m = \omega_m$$

The linear velocity of the DDMR in the LCS is the average of the linear velocities of the two wheels as [17-20]:
$$V_m = \frac{(V_{wr} + V_{wl})}{2} = r_w \frac{(\omega_{wr} + \omega_{wl})}{2} \quad (13)$$

And the angular velocity of the DDMR is
$$\omega_m = \frac{(V_{wr} - V_{wl})}{D} = r_w \frac{(\omega_{wr} - \omega_{wl})}{D} \quad (14)$$

Where $V_m$ is the longitudinal velocity of the center of mass, $\omega_m$ the angular velocity and heading of the robot, $V_{wr}$ and $V_{wl}$ are the longitudinal tire velocities of the right and left wheels respectively, $\omega_{wr}$ and $\omega_{wl}$ are the angular tire velocities of the right and left wheels respectively, and $r_w$ is the nominal radius of the tire.

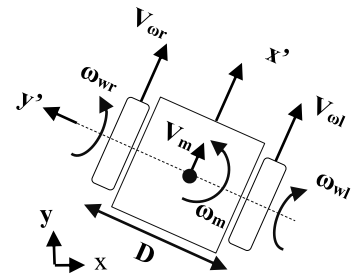

Fig.3 The differential drive mobile robot

In the work done by [13], the dynamical equations of the motor-wheels were presented in details. The state space representation of the overall motor and wheel dynamics can be summarized as follows (for the right wheel):
$$J_{eq} n \dot{\omega}_{wr} = -B_{eq} n \omega_{wr} + k_t i_{ar} - \tau'_{lr}$$
$$L_a \frac{di_{ar}}{dt} = -k_b n \omega_{wr} - R_a i_{ar} + v_{ar}$$
$$\tau'_{lr} = \tau_{rext}/n$$

where $v_{ar}$ and $v_{al}$ are the input voltages applied to the right and left motors respectively, $i_{ar}$ and $i_{al}$ are the armature current of the right and left motors respectively, $\tau'_{lr}$ and $\tau'_{ll}$ are the right and left motor developed torques, $k_b$ is equal to the voltage constant, $k_t$ is the torque constant, $R_a$ is the electric resistance constant, $L_a$ is the electric self-inductance constant, the total equivalent inertia, $J_{eq}$ and total equivalent damping, $B_{eq}$ at the armature of the combined motor rotor, gearbox, and wheel, $n$ is the gearbox ratio, and $\tau_{rext}$ and $\tau_{lext}$ are the external torque applied at the wheel side for the right land left wheels, respectively.

Let $x_1 = \omega_{wr}$, $x_2 = i_{ar}$, $d = \tau'_{lr}$, and $u = v_{ar}$
Then,
$$\dot{x}_1 = -\frac{B_{eq}}{J_{eq}}x_1 + \frac{k_t}{J_{eq}n}x_2 - \frac{1}{J_{eq}n}d$$
$$\dot{x}_2 = -\frac{k_b n}{L_a}x_1 - \frac{R_a}{L_a}x_2 + \frac{1}{L_a}u,$$
Let $b_1 = -\frac{1}{J_{eq}n}$, $b_2 = \frac{1}{L_a}$, $f_1(x_1, x_2) = -\frac{B_{eq}}{J_{eq}}x_1 + \frac{k_t}{J_{eq}n}x_2$, and $f_2(x_1, x_2) = -\frac{k_b n}{L_a}$

It can be seen that the simplified model with the mismatched uncertainties and external disturbances of the DDMR exactly fits the state-space formulation given in (7). According to theorem 1, the state-space model with mismatch uncertainties can be transformed into ADRC canonical form with $\hat{b} = \frac{1}{L_a}\frac{k_t}{J_{eq}n}$ for the motor wheel model.

## IV. NUMERICAL SIMULATION

In fig. 4, the kinematic model of the DDMR driven by a PMDC motors and controlled by our proposed IADRC is numerically simulated using MATLAB® /SIMULINK environment. The numerical simulations are done by using Matlab® ODE45 solver for the models with continuous states. This Runge-Kutta ODE45 solver is a fifth-order method that performs a fourth-order estimate of the error.

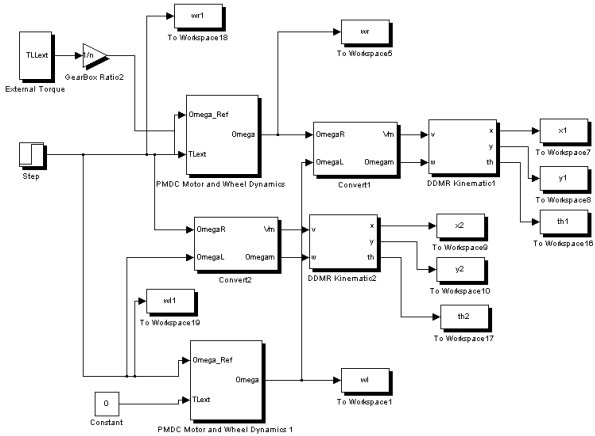

Fig.4 The Simulink® block diagram of the DDMR kinematics and the PMDC motor controlled by the IADRC.

The values of the parameters for PMDC motor are $R_a$=0.1557, $L_a$=0.82, $K_b$=1.185, $K_t$=1.1882, $n$=3.0, $J_{eq}$=0.2752, and $B_{eq}$=0.3922. The DDMR used in the simulation is assumed to have the following parameters: $D$=0.40, and $r_w$=0.075. The parameters of the classical ADRC controller are $\delta_1$=0.4620, $\delta_2$=0.24807, $\alpha_1$=0.1726, $\alpha_2$=0.8730, $\beta_1$=30.4, $\beta_2$=523.4, $\beta_3$=2970.8, and $R$=100.

The parameters of the proposed INLSEF are :$k_{11}$=144.6642, $k_{12}$=8.0475, $k_{21}$=25.5574, $k_{22}$=4.8814, $k_3$=0.5308, $\delta$=3.8831, $\mu_1$=44.3160, $\mu_2$=48.8179, $\mu_3$=26.1493, $\alpha_1$=0.9675, $\alpha_2$=1.4487,and $\alpha_3$=3.5032.

The ITD proposed in this work has the following set of parameters: $\alpha$= 0.4968, $\beta$=2.1555, $\gamma$=11.9554, $R$=16.8199. The parameters $K_\alpha$=0.6265, $\alpha$=0.8433, $K_\beta$=0.5878, $\beta$=0.04078, $\beta_0$=30.4, $\beta_1$=513.4, $\beta_2$=1570.8 represent the coefficients of the SMESO used in this work.

The DDMR is tested by applying a reference angular velocities for both wheels equal to 1 rad/s at t=0 and for t=100 sec. To investigate the performance of the proposed IADRC an external torque act as a constant disturbance is applied to the right wheel during the simulation at t=30 and removed after 20 sec. Fig. 5 shows the applied external disturbance. Fig. 6, shows the transient response of the controlled PMDC motor for the right wheel when both the ADRC and the IADRC have been applied. The figure shows an improvement in the response of the system with before, and during the presence of the applied disturbance when the IADRC has been adopted, this behavior can be noticed in fig. 6-d.

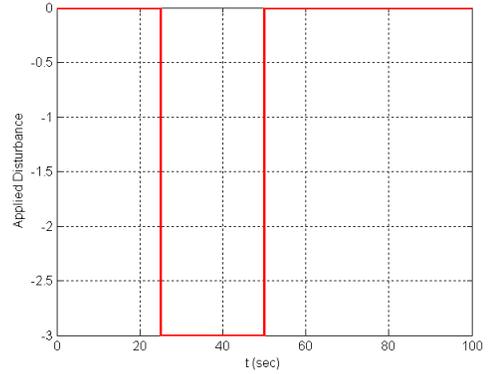

Fig.5. The applied external torque.

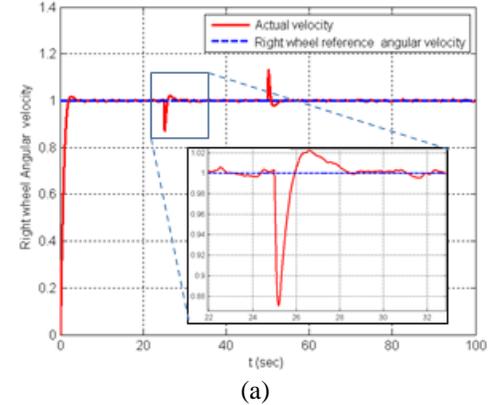

(a)

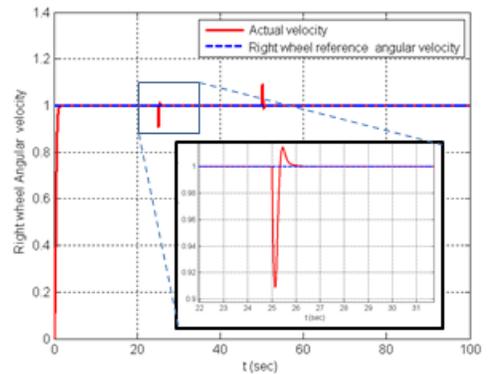

(b)

Fig.6. The simulation results, (a) The angular velocity of the right wheel using ADRC, (b) The angular velocity of left wheel IADRC

The orientation error $e_\theta$ associated with the tested case is reduced intensely due to the effectiveness of the proposed technique (see fig. 7). Note that, $e_\theta = \theta_{ref} - \theta_{actual}$ where $\theta_{ref}$ is the orientation of the reference trajectory and $\theta_{actual}$ is the actual orientation. It can be seen that the IADRC produces an error signal with less overshoot ($3.4 \times 10^{-3}$) than in the ADRC scheme ($10.5 \times 10^{-3}$). Also, the IADRC shows a faster convergence for the error signal. This is because of the proposed nonlinearities in the INLSEF controller which strongly and quickly damps the error signals.

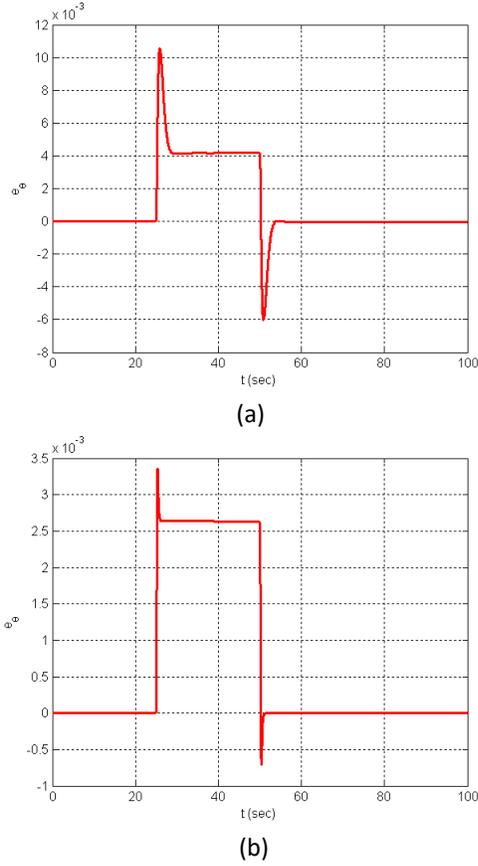

Fig.7. The simulation results, (a) the DDMR orientation error in case of ADRC, (b) the DDMR orientation error in case of IADRC.

The chattering phenomenon found in the estimated total disturbances produced by the LESO of the conventional ADRC for both wheels $D_r$ and $D_l$ is extremely reduced by using the SMESO of the proposed IADRC. The same is in the control signals that drive the two wheels $u_r$ and $u_l$ (see fig. 8), where a very smooth control signal has been obtained on account of a little increase in the overshot (compare fig. 8-a and fig. 8-b).

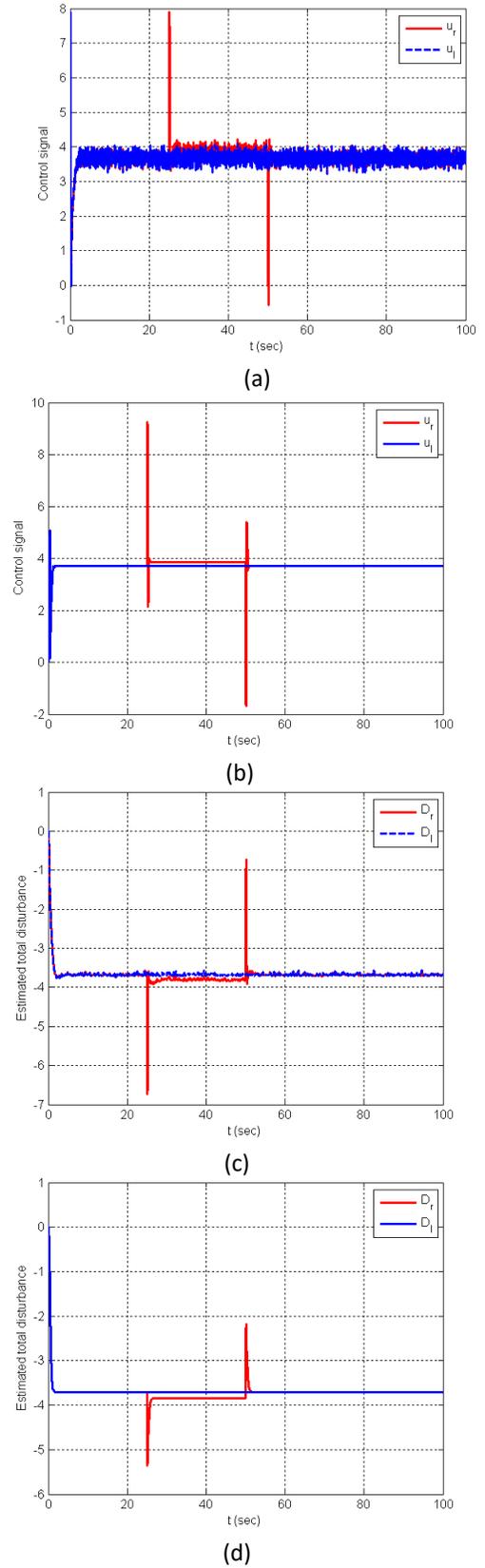

Fig.8 The simulation results, (a) the control signals generated by the ADRC, (b) the control signals generated by the IADRC, (c) the estimated input equivalent total disturbance from the LESO, (d) the estimated input equivalent total disturbance from the SMESO.

The results are collected based on evaluating several indices listed in tables I and II. The indices reflect the performance of the adaptive improved active disturbance rejection control. The results are classified into kinematics and dynamics performance indices.

TABLE I. THE DDMR KINEMATICS INDICES

| Performance Index | Controller | |
|---|---|---|
| | *ADRC* | *IADRC* |
| $OPI_x$ | 0.0010884970 | 0.0005257305 |
| $OPI_y$ | 0.0016112239 | 0.0007447036 |
| $OPI_\theta$ | 0.0000059780 | 0.0000017459 |

TABLE II. THE PERFORMANCE INDICES OF THE BOTH MOTORS

| Wheel | Performance Index | Controller | |
|---|---|---|---|
| | | *ADRC* | *IADRC* |
| Right | ITAE | 13.302889 | 1.780254 |
| | ISU | 1372.090423 | 1407.300305 |
| Left | ITAE | 6.919226 | 0.146694 |
| | ISU | 1343.542226 | 1372.124019 |

Where

$$OPI_x = \frac{1}{N}\sum(x_{ref} - x_{actual})^2,$$
$$OPI_y = \frac{1}{N}\sum(y_{ref} - y_{actual})^2,$$
$$OPI_\theta = \frac{1}{N}\sum(\theta_{ref} - \theta_{actual})^2$$
$$ITAE = \sum t|\omega_{ref} - \omega_{actual}|\,dt$$
$$ISU = \sum u^2\,dt$$

A major improvement in the kinematic indices for the IADRC against the conventional ADRC, where The $OPI_x$, $OPI_y$, and $OPI_\theta$ reduced by 51.7%, 53.78%, and 70.794% respectively. The simulations show that the ISU which represents the energy delivered to the PMDC motor has been increased by 2.566% with a noticeable improvement in the transient response (ITAE is reduced by 86.6175%). In addition, the chattering in the control signal caused by Han's conventional ADRC is almost eliminated by the proposed IADRC. Finally, the DDMR orientation error has been clearly reduced and swiftly decreases to zero.

## V. CONCLUSION

An improved nonlinear ADRC controller was developed for a DDMR to achieve accurate speed tracking in the presence of high external torque disturbance. The proposed IADRC with the SMESO generates an exact estimation of the states and the total disturbance. The proposed IADRC with the three parts, namely, the SMESO, the NLSEF, and the INTD provide a promising scheme to improve the capability of the conventional ADRC for disturbance estimation and rejection. Based on the simulation results, it can be concluded that the developed IADRC can effectively improve the accuracy and the speed of the PMDC motor of the DDMR under mismatched uncertainties and torque disturbance. The IADRC eliminates the chattering phenomenon, which is coherent in the conventional ADRC with little increase in the overshoot of the control signal at the instance of the disturbance occurrence. The future directions for our proposed IADRC are extending the applications to include consensus multi-agent systems. The first step will be to design a control system for every local agent for consensus disturbance rejection. Secondly, analyzing the design for network-connected multi-input linear or nonlinear systems using relative state information of the subsystems in the neighborhood. The configuration of the consensus multi-agent system can be in leaderless and leader-follower consensus set-ups under some common assumptions of the network connections.